\input harvmac
\noblackbox

\input epsf

\def\tilde{\widetilde}
\newcount\figno
\figno=0
\def\fig#1#2#3{
\par\begingroup\parindent=0pt\leftskip=1cm\rightskip=1cm\parindent=0pt
\baselineskip=11pt
\global\advance\figno by 1
\midinsert
\epsfxsize=#3
\centerline{\epsfbox{#2}}
\vskip 12pt
{\bf Fig.\ \the\figno: } #1\par
\endinsert\endgroup\par
}
\def\figlabel#1{\xdef#1{\the\figno}}
\def\encadremath#1{\vbox{\hrule\hbox{\vrule\kern8pt\vbox{\kern8pt
\hbox{$\displaystyle #1$}\kern8pt}
\kern8pt\vrule}\hrule}}
\def\apm{{\alpha^{\prime}}}


\font\cmss=cmss10
\font\cmsss=cmss10 at 7pt
\def\rlx{\relax\leavevmode}
\def\inbar{\vrule height1.5ex width.4pt depth0pt}
\def\IC{\relax\,\hbox{$\inbar\kern-.3em{\rm C}$}}
\def\IN{\relax{\rm I\kern-.18em N}}
\def\IP{\relax{\rm I\kern-.18em P}}
\def\ZZ{\rlx\leavevmode\ifmmode\mathchoice{\hbox{\cmss Z\kern-.4em Z}}
 {\hbox{\cmss Z\kern-.4em Z}}{\lower.9pt\hbox{\cmsss Z\kern-.36em Z}}
 {\lower1.2pt\hbox{\cmsss Z\kern-.36em Z}}\else{\cmss Z\kern-.4em
 Z}\fi}
\def\IZ{\relax\ifmmode\mathchoice
{\hbox{\cmss Z\kern-.4em Z}}{\hbox{\cmss Z\kern-.4em Z}}
{\lower.9pt\hbox{\cmsss Z\kern-.4em Z}}
{\lower1.2pt\hbox{\cmsss Z\kern-.4em Z}}\else{\cmss Z\kern-.4em
Z}\fi}
\def\IZ{\relax\ifmmode\mathchoice
{\hbox{\cmss Z\kern-.4em Z}}{\hbox{\cmss Z\kern-.4em Z}}
{\lower.9pt\hbox{\cmsss Z\kern-.4em Z}}
{\lower1.2pt\hbox{\cmsss Z\kern-.4em Z}}\else{\cmss Z\kern-.4em
Z}\fi}

\def\narrowplus{\kern -.04truein + \kern -.03truein}
\def\narrowminus{- \kern -.04truein}
\def\narrowminussub{\kern -.02truein - \kern -.01truein}

\def\half{{1\over 2}}

\def\b{{\beta}}

\def\a{{\alpha}}

\def\D{{\Delta}}
\def\m{{\mu}}
\def\n{{\nu}}
\def\ep{{\epsilon}}
\def\d{{\delta}}

\def\ph{{\phi}}
\def\t{{\theta}}
\def\T{{\Theta}}
\def\l{{\lambda}}

\def\r{{\rightarrow}}

\def\frac#1#2{{#1\over #2}}

\def\CN{{\cal N}}

\def\IZ{\relax\ifmmode\mathchoice
{\hbox{\cmss Z\kern-.4em Z}}{\hbox{\cmss Z\kern-.4em Z}}
{\lower.9pt\hbox{\cmsss Z\kern-.4em Z}}
{\lower1.2pt\hbox{\cmsss Z\kern-.4em Z}}\else{\cmss Z\kern-.4em
Z}\fi}
\def\IB{\relax{\rm I\kern-.18em B}}
\def\IC{{\relax\hbox{$\inbar\kern-.3em{\rm C}$}}}
\def\ID{\relax{\rm I\kern-.18em D}}
\def\IE{\relax{\rm I\kern-.18em E}}
\def\IF{\relax{\rm I\kern-.18em F}}
\def\IG{\relax\hbox{$\inbar\kern-.3em{\rm G}$}}
\def\IGa{\relax\hbox{${\rm I}\kern-.18em\Gamma$}}
\def\IH{\relax{\rm I\kern-.18em H}}
\def\II{\relax{\rm I\kern-.18em I}}
\def\IK{\relax{\rm I\kern-.18em K}}
\def\IP{\relax{\rm I\kern-.18em P}}

\def\p{\partial}

\font\cmss=cmss10 \font\cmsss=cmss10 at 7pt
\def\IR{\relax{\rm I\kern-.18em R}}

%

%
%
\def\eqnn#1{\xdef #1{(\secsym\the\meqno)}\writedef{#1\leftbracket#1}%
\global\advance\meqno by1\wrlabeL#1}
\def\eqna#1{\xdef #1##1{\hbox{$(\secsym\the\meqno##1)$}}
\writedef{#1\numbersign1\leftbracket#1{\numbersign1}}%
\global\advance\meqno by1\wrlabeL{#1$\{\}$}}
\def\eqn#1#2{\xdef #1{(\secsym\the\meqno)}\writedef{#1\leftbracket#1}%
\global\advance\meqno by1$$#2\eqno#1\eqlabeL#1$$}

\lref\sstone{N. Seiberg, L. Susskind and N. Toumbas, {\it The Teleological
Behavior of Rigid Regge Rods}, hep-th/0005015. }
\lref\ssttwo{N. Seiberg, L. Susskind and N. Toumbas, 
{\it Strings in background 
electric field, space / time noncommutativity  
and a new noncritical string theory},
hep-th/0005040.}
\lref\msv{ Shiraz Minwalla, Mark Van Raamsdonk, Nathan Seiberg,
{ \it Noncommutative Perturbative Dynamics }, hep-th/9912072.}
\lref\bach{C. Bachas and M. Porrati, {\it
Pair Creation of Open Strings in an Electric Field},
Phys.Lett.{\bf B296}:77-84,1992.}
\lref\nest{V.V.Nesterenko, {\it 
The Dynamics of Open Strings in a Background Electromagnetic Field}, 
Int. J. Mod. Phys. {\bf A4} (1989) 2627-2652.}
\lref\russo{Chong-Sun Chu, Rodolfo Russo and Stefano Sciuto,
{\it Multiloop String Amplitudes with B-Field and Noncommutative QFT},
hep-th/0004183.}
\lref\call{C.G. Callan, C. Lovelace, C.R. Nappi, S.A. Yost,
{\it Open Strings in Background gauge Fields}, 
Nucl.Phys.{ \bf B288} 525,1985. }
\lref\tsyt{E.S. Fradkin, A.A. Tseytlin,
{ \it Nonlinear Electrodynamics from Quantized Strings} 
Phys.Lett.{\bf B163} 123,1985.}
\lref\mich{ Hong Liu and Jeremy Michelson,
{\it Stretched Strings in Noncommutative Field Theory},
 hep-th/0004013.   }
\lref\burg{C.P.Burgess, {\it 
Open String Instability in Background electric Fields}, 
Nuc. Phys. {\bf B294}427-444, 1987.}
\lref\cds{Alain Connes, Michael R. Douglas and Albert Schwarz,
{\it  Noncommutative Geometry and Matrix Theory: Compactification on Tori},
hep-th/9711162, JHEP 9802 (1998) 003.}
\lref\sw{N. Seiberg and E. Witten
{\it String Theory and Noncommutative Geometry}, hep-th/9912072.}
\lref\whre{ A.A. Tseytlin, {\it 
Born-Infeld action, supersymmetry and string theory}
hep-th/9908105, and references therein.}
\lref\gomis{Jaume Gomis, Matthew Kleban, Thomas Mehen, Mukund Rangamani and
Stephen Shenker, {\it Noncommutative Gauge Dynamics From The String
Worldsheet},hep-th/0003215. }
\lref\kiem{Youngjai Kiem and Sangmin Lee, 
  {\it UV/IR Mixing in Noncommutative Field Theory via Open String Loops},
  hep-th/0003145. }
\lref\chu{Adel Bilal, Chong-Sun Chu, Rodolfo Russo, 
{\it String Theory and Noncommutative Field Theories at One Loop},
hep-th/0003180.}
\lref\andreev{Oleg Andreev and Harald Dorn, 
{\it Diagrams of Noncommutative Phi-Three Theory from String Theory },
hep-th/0003113.}
\lref\oleg{O.~Andreev,
{\it A note on open strings in the presence of constant B-field},
Phys.\ Lett.\  {\bf B481}, 125 (2000)
hep-th/0001118.}
\lref\wittenopen{E.~Witten,
{\it Noncommutative Geometry And String Field Theory},
Nucl.\ Phys.\  {\bf B268}, 253 (1986).
}
\lref\arvind{ Arvind Rajaraman and  Moshe Rozali,
{\it Noncommutative Gauge Theory, Divergences and Closed Strings},
hep-th/0003227.}
\lref\mr{J. Maldacena and J. Russo, {\it The Large N limit of
non-commutative
gauge theories}, hep-th/9908134, JHEP 9909:025,1999.}
\lref\schom{V. Schomerus, {\it D-branes and Deformation Quantization},
hep-th/9903205, JHEP 9906 (1999) 030.}
\lref\sv{M. van Raamsdonk and N. Seiberg,
{\it Comments on Noncommutative Perturbative Dynamics},JHEP 0003 (2000) 035}
\lref\ih{A. Hashimoto and N. Itzhaki, {\it Non-Commutative 
Yang-Mills and the AdS/CFT Correspondence}, 
Phys.Lett. B465 (1999) 142}
\lref\rfilk{T. Filk, {\it Divergences in a Field Theory on Quantum Space}, 
 Phys.Lett.{ \bf B376}:53-58,1996.} 
\lref\sunny{N. Itzhaki, private communication.}
\lref\barbon{ 
J.~L.~Barbon and E.~Rabinovici,
{\it Stringy fuzziness as the custodian of time-space noncommutativity},
hep-th/0005073.}
\lref\sethi{O.~J.~Ganor, G.~Rajesh and S.~Sethi,
{\it Duality and non-commutative gauge theory},
hep-th/0005046.}
\lref\mehen{ J.~Gomis and T.~Mehen,
{\it Space-time noncommutative field theories and unitarity},
hep-th/0005129.}

\Title{\vbox{\baselineskip12pt\hbox{hep-th/0005048}\hbox{}
\hbox{}}}{S-Duality and Noncommutative Gauge Theory }

\centerline{Rajesh Gopakumar, Juan Maldacena, 
Shiraz Minwalla and  Andrew Strominger}
\bigskip\centerline{Jefferson Physical Laboratory,}
\centerline{Harvard University,} \centerline{Cambridge, MA 02138}

\vskip .3in \centerline{\bf Abstract} { It is conjectured that
strongly coupled, spatially noncommutative $\CN=4$ Yang-Mills
theory has a dual description as a weakly coupled open string theory in a
near critical electric field, and that this dual theory is fully
decoupled from closed strings. Evidence for this conjecture
is given by the
absence of physical closed string poles in the non-planar one-loop
open string diagram. The open string theory can be viewed as
living in a geometry in which space and time coordinates do not
commute.  }
\smallskip
\Date{} 
\listtoc 
\writetoc

\newsec{Introduction}

Noncommutative field theories have a rich and fascinating
structure. The embedding of these theories into string theory
\cds\ suggests that this structure may be directly relevant to
understanding the inevitable breakdown of our 
familiar notions of space and time at short
distances in quantum gravity.

Investigations to date have largely concentrated on 
theories with purely spatial noncommutativity (see however 
\sstone). While such theories are interestingly nonlocal in space, 
they are local in time, admitting familiar
notions like that of the Hamiltonian and a quantum state. 
Noncommutativity of  a time-like coordinate should have even more far-reaching
consequences, and it is natural to ask whether or not such
theories exist.

\lref\gkp{S.~Gukov, I.~R.~Klebanov and A.~M.~Polyakov,
{\it Dynamics of (n,1) strings},
Phys.\ Lett.\  {\bf B423}, 64 (1998)
[hep-th/9711112].}
\lref\verl{H.~Verlinde,
{\it A matrix string interpretation of the large N loop equation},
hep-th/9705029.}

In this paper we give one answer to this question by asking
another: What is the strong coupling dual of NCYM
(spatially-noncommutative $\CN=4$ Yang-Mills)? This 
question can be addressed
in the description of NCYM as a
scaling limit of three-branes with a $B$ field in $IIB$ string
theory \sw. $IIB$ $S$-duality induces an $S$-duality on the
NCYM theory, mapping the strongly coupled NCYM theory to a weakly
coupled open string theory\foot{The low energy sector of 
the open string theory is  ordinary $\CN=4$ YM, and the induced duality
reduces to the standard $S$-duality. }.
 This open string theory can be viewed
either as living in a near critical electric field\foot{The existence of 
a scaling theory at near
critical electric fields, and its relevance to temporal
noncommutativity was emphasized to us by N. 
Seiberg, L Susskind and N. Toumbas (private communications).
The scaling to the critical electric field was also considered in 
\gkp, \verl.}~ \foot{The
critical value of the electric field arises when the force pulling
apart the charges at either end of the string just balances the
string tension, so that the string is effectively tensionless
\refs{\call \burg \bach -\nest, \tsyt}. 
Beyond this value the spectrum contains a tachyon and the
vacuum is unstable.} , or in a space-time with noncommuting space and
time coordinates. A precise statement of the spacetime noncommutativity
in this theory is that the temporal
zero mode $X^0$ on the open string worldsheet does not commute
with the spatial zero modes. The scale associated with this
noncommutativity is the same as the effective open string scale.
Thus the effects of the noncommutativity are 
inextricably tied up with the usual stringy nonlocalities.

Since the closed string sector of the $IIB$
theory is decoupled in the scaling limit, the dual open string
theory does not have a closed string sector.
The appearance of an open string theory without a closed string
sector is striking. Ordinarily closed string poles appear in open
string loop diagrams, and unitarity then requires the addition of
asymptotic closed string states. In order to better understand
this point we analyze (following \refs{\andreev 
\kiem \gomis \mich \chu -\russo}) the nonplanar one loop
open string diagram for the bosonic case. 
We find that the temporally noncommutative
phases lead to a precise cancellation of all the closed string
poles, in accord with our expectations. This cancellation in fact occurs
for branes of any dimension, indicating the existence of 
a family of non-commutative open string theories.  

This paper is organized as follows. In section 2 we derive the 
$S$-dual of NCYM, which we refer to as NCOS (noncommutative 
open strings), by embedding in string theory. In section 3 we
 show that it is a decoupled open string theory with a 
near-critical electric field. 
In section 4 we give evidence at the one loop 
level for the decoupling of closed strings by computing  
the non-planar annulus for bosonic string theory with two incoming and two
outgoing tachyons. Section 5 contains a preliminary analysis of the 
general higher loop diagram; no obvious 
closed string singularities are found. 
In section 6 we make some comments regarding the supergravity duals of
our open string theory.
We conclude with some discussion in section 7.  For simplicity we concentrate 
on the $U(1)$ theories but our results generalize easily to $U(N)$. 

Related work will appear in \ssttwo.  

\newsec{Inducing S-Duality}

The Olive Montonen dual of ordinary $\CN=4$ SYM may be deduced as a
consequence of the $S$ duality of IIB theory
in the presence of D3-branes in the zero slope limit.
In this section we will determine the Olive Montonen dual of
spatially noncommutative $\CN=4$ SYM, using the $S$ duality of
IIB theory in flat space in the presence of D3-branes and a
background $B_{\m\n}$ field, together with the modified zero slope limit
\sw\ .

Consider a D3-brane, extended in the $0,1,2,3$ directions,
in a background geometry
\eqn\frc{\eqalign{g^{\prime}_{\m\n}&=\eta_{\m\n}, \  \  
g^{\prime}_{ij}=\apm^2k_1 \d_{ij}, \  \  g^{\prime}_{MN}= \d_{MN}, 
\cr B^{\prime}_{ij}&=-B \ep_{ij},
\  \  g^{\prime}_{str}=\apm k_2. }}
in the limit $\apm \r 0$, keeping $k_1,k_2,B$ fixed 
(we will refer to this as the NCYM limit).
Here $\m,\n =0,1$ with $i,j=2,3$ and $M,N=4,...9$. (We will reserve 
unprimed notation for the S-dual variables to be introduced in the 
next sub-section.)
It was shown in \sw\ that the decoupled theory on the
brane is noncommutative $U(1)$ SYM propagating on a four dimensional
space with (open string) metric (we use the conventions of \sw)
$G^{\prime}_{\m\n}=\eta_{\m\n},~~ G^{\prime}_{ij}={( 2 \pi B)^2 \over k_1} \d_{ij}$,
noncommutativity parameter $\t^{\prime ij}={\ep^{ij} \over B}$,
and gauge coupling $g_{YM}^2=2\pi G_o^{'2}$, where
$G_o^{\prime 
2}={k_2 B\over k_1}$.  
In order to obtain noncommutative field theory  propagating on a space
with unit metric we choose $k_1=(2\pi B)^2$.
In terms of the field theory couplings $\t^{'}$ and $G'_o$,
$B={1 \over \t^{'}}$ and $k_2={(2 \pi)^2 G_o^{'2} \over \t^{'}}$.

In order to obtain a weakly-coupled dual
description of the noncommutative gauge
theory at large $G'_o$
we will consider the NCYM limit described above
in an $S$-dual picture. Before describing this in detail
we note that the $S$-dual version has two potentially unpleasant features:

\item{a.}
It seems to involve branes in the presence of an
an RR 2 form potential (the $S$-dual of $B'_{ij}$).

\item{b.} The S-dual of the NCYM limit takes the closed string coupling $g_{str}$ to
infinity, seeming to indicate that any description of brane dynamics
obtained in this picture will be strongly rather than weakly coupled, 
independent of $G_o$.

These difficulties may both be circumvented. In order to avoid having to
deal with RR fields, we gauge away
the constant bulk NS-NS potential before performing the S-duality.
This gauge transformation induces a magnetic field $F'_{23}=B$ on the
the D3-branes, which is converted into
an electric field by the $S$-duality; in fact an
electric field that approaches its critical value in the scaling
limit.
This electric field may in turn
be gauged into a constant background
NS-NS two form potential $B_{01}=F_{01}$ in the bulk.
But, in such a background, the open string coupling
that governs the strength of interactions between brane modes
is not directly related to the closed string coupling. It turns
out that the open string coupling in this background is
 $G_o= {1 \over G'_o}$, i.e. it is the inverse of the original open string
coupling, and therefore remains finite despite the fact that
$g_{str} \to \infty$.   Thus at large $G'_o$, the effective description
is a weakly coupled noncommutative open string theory, with noncommutativity in the
time direction!

We now consider this limit in more detail. We could consider any finite
number of branes, $N$, but we will mostly 
stick to the case $N=1$ for simplicity.

\subsec{Born-Infeld $S$-Duality}

$S$-duality transforms a constant magnetic field on the three-brane
to a constant electric field. Constant fields 
on a single D3-brane are governed by the Born-Infeld action
\eqn\bi{S_{BI}={1 \over (2\pi)^3\apm^2 g_{str}} \int d^4x 
\sqrt{ - \det(g_{\m\n}-2\pi\apm F_{\m\n})}.}
The action of $S$ duality on $S_{BI}$ will be reviewed in this
subsection (See \whre\ ).
Consider a gauge theory on a torus. The flux of the magnetic field on
any nontrivial two cycle of the torus is integrally quantized, and so must,
under electromagnetic $S$-duality, map to a quantized electric flux.
Recall why electric flux on a torus is quantized.
The constant piece (zero momentum mode) of a gauge field in flat infinite
space is physically unmeasurable, as it can be gauged away. This is not
true, however, on a torus, as the Wilson line $e^{i\int A.dx}$ over any
nontrivial cycle of the torus is a gauge invariant observable, implying
that the zero momentum piece of the gauge field $A_{i}$ is a periodic
physical `coordinate', with period ${2 \pi \over L_{i}}$
($L_{i}$ is the size  of the  $i^{th}$ spatial direction).
Consequently, the momentum conjugate to the zero mode of $A_{i}$ is
quantized in integral units of $L_{i}$.
This quantized momentum is the electric flux that is  interchanged
with the quantized magnetic flux under $S$ duality.

In order to work out the expression for the
quantized electric flux, consider the theory \bi\
on a rectangular torus, with spatial coordinate radii $L_1, L_2, L_3$.
We are interested in  background
field configurations in which $F_{01}$ is nonzero and constant, but
$F_{ij}$ is zero. Since $\dot{A}_1$ appears in the Lagrangian only
through $F_{01}$, it is sufficient, for the purposes of computing
canonical momenta in such backgrounds, to
set $F_{ij}$ to zero in the Lagrangian.
For a diagonal metric the Born Infeld action
simplifies to (recall $g^{00}$ is negative)
\eqn\bis{S={1 \over (2\pi)^3\apm^2 g_{str}} \int d^4x
\sqrt{-g}\sqrt{ 1+(2\pi\apm)^{2}
g^{11}g^{00}F_{01}^2}.}
Thus, for constant $F_{01}$,
 the momentum conjugate to
$A_1$ is
\eqn\conmom{P^{1}=N L_1={1 \over 2\pi g_{str}} L_1 L_2 L_3
\sqrt{-g}{ g^{11} g^{00} F_{01} \over
\sqrt{ 1+(2\pi\apm)^{2}
g^{11}g^{00} F_{01}^2}}. }

Thus the constant $F'_{23}$ background of the spatially noncommutative
theory maps, under
$S$ duality, to a background with constant $F_{01}$, whose value
is given by the solutions to the equations
\eqn\valelfield{
{\sqrt{-g} \over g_{str}}{ g^{11} g^{00} F_{01} \over
\sqrt{ 1+(2\pi\apm)^{2}
g^{11}g^{00} F_{01}^2}} =F'_{23}={1 \over \t^{'}} }
where $g_{\m\n}$ and $F_{\m\n}$ are the background
metric and field strength in the $S$ dual description.
In terms of the critical value of the electric field
\eqn\crv{F_{01}^{crit}={\sqrt{-g_{00}g_{11}} \over 2 \pi \apm}}
one finds
\eqn\drf{F_{01}={F_{01}^{crit}\over \sqrt{1+g_{22}g_{33}
({\t^{'}\over 2 \pi  \apm g_{str}})^2}}.}
\subsec{The Scaling Limits}

Consider IIB theory with a D3-brane 
in the presence of a background NS-NS 2-form 
potential, $B_{\m\n}$. 
Prior to any scaling limit, an open string  metric $\tilde{G}^{AB}$
(the symbol $G^{AB}$ will be reserved for a rescaled open string metric
defined below)
and a non-commutativity parameter $\Theta$
can be deduced from disk correlators on the open string worldsheet boundaries
\eqn\rfz{X^A(0)X^B(\tau)= -\apm \tilde{G}^{AB}\ln(\tau)^2+{i\over 2}
\Theta^{AB}\epsilon(\tau),~~~~~
A,B=0,1,2,3.}
The open string coupling $G_o$ is similarly read off from the coefficient
of the gauge theory action. These are related to closed string quantities
by the formulae \sw
\eqn\gya{\eqalign{2\pi \apm\tilde{G}^{AB}+\T^{AB}&=\bigl(2\pi\apm)({1 \over g+2\pi
\apm B}\bigr)^{AB},\cr
              G_o^2&=g_{str}{{\rm det}^{\half}(g+2\pi
\apm B) \over {\rm det}^{\half}(g)}.}} 

As discussed above, in the NCYM limit,
$\apm \to 0$ while the open string metric $G^{'AB}$, open string coupling 
$G_o'$ and the (spatial) non-commutativity matrix $\T^{'A B}$
are held fixed. We would now like to study this scaling limit 
in the S-dual description of Type $IIB$ theory. We will call this the NCOS limit.
Under an S-Duality, the type
$IIB$ closed string backgrounds transform in the usual fashion, 
$g^{'}_{str}={1 \over g_{str}}$, $g^{\prime}_{\m\n}={g_{\m\n}\over
g_{str}}$ ($\apm$ is unchanged). The 
associated open string quantities may then be read from their definitions
in \gya. The results, in the limit $\apm \r 0$,  are
summarized in the following table:
\medskip
\centerline{TABLE 1}
\medskip
\centerline{\vbox{\offinterlineskip
\hrule
\halign{&\vrule#&
        \strut\quad#\quad\cr
height3pt&\omit&&\omit&\cr &The NCYM Limit \hfil && The $S$-Dual
NCOS Limit \hfil & \cr height3pt&\omit&&\omit&\cr \noalign{\hrule}
height3pt&\omit&&\omit&\cr & $g^{'}_{\m\n}=\eta_{\m\n}$ \hfil &&
 $g_{\m\n }={\t G_o^4 \over 2\pi\apm} \eta_{\m\n}$\hfil &\cr
&  $g^{'}_{ij }={ (2\pi\apm)^2 \over \t^{'2}}\d_{ij}$ \hfil &&
$g_{ij}={2\pi\apm \over \t }\d_{ij}$  \hfil &\cr
& $B'_{\m\n}=F'_{\m\n}=0$  \hfil &&
$B_{\m\n}=F_{\m\n}=
F^{crit}_{\m\n}\left(1-\half \left( 
{2\pi\apm \over \t G_o^2}\right)^2 \right)$ \hfil & \cr
& $B'_{ij}=F'_{ij}=-{1 \over \t^{'}}\ep_{ij}$  \hfil &&
$B_{ij}=F_{ij}=0$ \hfil & \cr
& $g^{'}_{str}=G_o^{'2}{ 2\pi\apm \over \t^{'}}$ \hfil
&& $g_{str}={\t^{'} \over G_o^{'2} 2\pi\apm}={G_o^4\t \over 2\pi\apm}$ \hfil & \cr
& $G^{'AB}=\eta^{AB}$  \hfil &&
${\apm\over \apm_{eff}}\tilde{G}^{AB}\equiv G^{AB}=
\eta^{AB}$ \hfil & \cr
& $G^{'MN}=g^{'MN}=\d^{MN}$ \hfil &&
$G^{MN}=g^{MN}={2\pi\apm\over \t G_o^4}\d^{MN}$ \hfil & \cr
& $\T^{'\m\n}=0$ \hfil &&
$\T^{\m\n}=-\t^{'} G_o^{'2} \ep^{\m\n}=-\t\ep^{\m\n} $ \hfil & \cr
& $\T^{'ij}={-\t^{'} \ep^{ij}}$ \hfil &&
$\T^{ij}=0$ \hfil & \cr
&$ G_o^{'} =G_o^{'} $ \hfil &&
$G_o={1 \over G'_o}$ \hfil & \cr
&$ \apm =\apm $ \hfil &&
$\apm_{eff}={\t \over 2 \pi}$ \hfil & \cr
height3pt&\omit&&\omit&\cr}
\hrule} }
\noindent  

Here $$\m ,\n=0,1,~~~~~i,j=2,3,~~~~~A,B=0,1,2,3, ~~~~~ M,N=4,5,6,7,8,9.$$ 
In Table 1 we have expressed all open and 
closed string quantities as functions of 
$\t$ and $G_o$, the noncommutativity parameter and open string 
coupling in the (S-dual) NCOS theory. We have also defined the quantities,
$\apm_{eff}$ the effective open string scale and 
the rescaled open string metric $G^{AB}={\apm \over
\apm_{eff}}\tilde{G}^{AB}$ of the NCOS theory.  

\noindent Note that
 
\item{1.}  In the limit $\apm \r 0$, 
the electric field $F_{01}$ of the NCOS theory attains its critical value 
\eqn\fcfd{F^{crit}_{01}= {\t G_o^4 \over (2 \pi \apm)^2}.}
\item{2.} The energy per unit coordinate length of 
an NCOS  open string stretched in the 1 direction is given by 
(recall that the ends of an open string are charged)
\eqn\epul{p_0={\ep_{01} \over 2\pi}\left(
{1 \over \apm}-2 \pi \ep^{01}F_{01} \right)\D x^1={1 \over 4 \pi \a'_{eff}}
\D x^1}
so these open strings have an effective  tension set by $\apm_{eff}$.
As a consequence, it will turn out that in the NCOS limit  
excited open string oscillator states are part of 
the decoupled theory on the brane in the NCOS limit, and 
that their mass scale is also set by  $\apm_{eff}$ .
\item{3.} The open string coupling
$G_o$ is the inverse of the gauge coupling $G_o'$ in the NCYM limit.

To summarize, strongly coupled 
spatially noncommutative Yang-Mills theory has an effective description
as a weakly coupled open string theory 
living on D3-branes, in the presence of a near critical electric field. 
The parameters of this open string theory are listed in Table 1. 
We will explore this theory in the rest of this paper.

\newsec{The Classical NCOS Theory}

\subsec{Spacetime Noncommutativity}

In the NCOS limit, open strings on the brane 
propagate in a background electric field. This results
in temporal noncommutativity,
in the sense that the open string zero modes obey
\eqn\ncmv{[X^\m,X^\n]=i\t \ep^{\m\n},} as may easily be seen
from \rfz. 

Disk diagrams in the NCOS theory are very simple.
As argued in \schom, \sw, open string correlation functions 
on the disk in the NCOS theory may be obtained from the equivalent
correlation functions in the theory without the electric field, by 
the addition of noncommutative phases in the $0,1$ directions
(and using the appropriate open string metric and coupling).  
Thus the classical action for open string modes 
in the NCOS limit may be obtained by turning all products in 
the usual open string classical action into star products.
In other words if we think about the open string field theory 
action $S = \int A QA + A *_w A *_w A $ there the $*_w$ product
is Witten's star product \wittenopen\ then the only change is
that we replace Witten's product by a modified product which 
just adds in the Moyal phases, and of course we replace $\alpha' 
\to \alpha'_{eff}$.

Since the effective string scale $\apm_{eff}$ is 
the same as that of non-commutativity $\t$,  
the noncommutative phases are non negligible only for energies of the order
of those of string oscillators.

\subsec{The Free Spectrum}

In this subsection we will argue that  the NCOS limit 
defines an open string theory on the 3-brane, as open string oscillators 
do not decouple in this limit. We will examine the spectrum in the free NCOS
theory and see that the effective scale is indeed set by $\apm_{eff}$.

We first consider the scaling limit in the  
NCYM picture.
Near the NCYM limit one has  weakly coupled closed strings
coupled gravitationally to open strings.
Open string excitations with
string frame energies obeying  $|g^{00}k_0^2| \ll  {1 \over \apm}$,
or equivalently Einstein frame energies obeying
$|g_{E}^{00}k_0^2| \ll m_p^2$ ,
decouple from the closed strings.
As $g^{00}=-1$, open string modes with $k_0 \ll { 1 \over \sqrt{\apm}}$
decouple from closed string modes.
The decoupled theory includes all brane excitations with energies that
obey this inequality, namely just the $\CN=4$ YM multiplet.

Now consider the same limit in the NCOS picture.
The argument  above ensures that open string modes with $k_0 \ll { 1 \over
\sqrt{\apm}}$ decouple from closed strings.  
However, the 
open string oscillator states in this picture  obey the
mass shell condition set by the open string metric in the RHS  of Table 1
\eqn\msc{{\apm_{eff} \over \apm} k_{A}G^{AB}k_{B}={N\over \apm}}
with $A,B=0,1,2,3$.
This implies
\eqn\mscp{k^2={N \over \apm_{eff}} \ll {1 \over \apm}}
with
\eqn\adf{\apm_{eff}={ \t \over 2\pi}}
in the limit $\apm \r 0$. Thus the decoupled theory on the brane includes
all open string oscillator states! The mass spectrum is exactly the usual
free spectrum on the three brane, except with
$\apm_{eff}$ replacing $\apm$.

\subsec{Worldsheet Correlators}

Nontrivial vertex operators are functions of 
tangential worldsheet  derivatives of $X^A$ and normal worldsheet
derivatives
of $X^M$. 
Correlation functions of such vertex operators may be computed given
the two point correlators of the free fields  $X^A$ restricted 
to the boundary of the world sheet, as well as the two point 
functions of the free fields $X^{M}$.

The boundary correlators of $X^A$ are finite in the limit $\apm \r 0$,
and are given by
\eqn\rfz{X^A(0)X^B(\tau)= -\apm_{eff}G^{AB}\ln(\tau)^2+{i\over 2}
\t^{AB}\epsilon(\tau),~~~~~
A,B=0,1,2,3  }

On the other hand, correlation functions involving the transverse  
directions $X^{M}$ are  derived from the sigma model 
\eqn\ssm{S={1\over 4\pi\apm}\int G_{MN}\p X^M\p X^N =
{\apm_{eff} G_0^4 \over 4 \pi \apm^2} 
\int \p X^{M} \p X^N \d_{MN}.} 
In terms of the rescaled fields 
$Y^{M} ={G_0^2 \apm_{eff} \over \apm}X^{M}$ 
\eqn\ssmm{S={1 \over 4 \pi \apm_{eff}} 
\int \p Y^M\p Y^N\d_{MN} .} 
The vertex operators representing physically normalized states 
are functions of the normal derivatives of $Y^{M}$. 

Thus all correlation functions of NCOS vertex 
operators on the disk will be the same as in usual open string theory
except that $\alpha' \to \alpha'_{eff}$ and we have extra non-commutative
phases appearing as in \sw . The open string coupling constant is 
$G_0$ and it is finite.

\newsec{The One Loop Diagram}

One loop open string graphs usually contain closed string poles, and 
unitarity then requires that closed strings be included as asymptotic
states. 
In this section we consider the nonplanar annulus diagram
in the NCOS limit, and show that it has no physical closed  
string poles. This demonstrates that
an on shell closed string cannot be produced
in collisions of open strings. 

Nonplanar diagrams for spatial $\T$ were computed in 
\refs{\andreev \kiem \gomis  \mich -\russo} --  we will follow \mich.
The nonplanar diagram for our case can be obtained  
by analytic
continuation. For simplicity consider the case of two initial and
two final open string tachyon vertex operators
$V_T=G_oe^{ik_AX^A}$ in the bosonic string with incoming momenta
$k_1,k_2$ and outgoing momenta $k_3,k_4$. Then we get for 
a D-3 brane in bosonic string theory   (Eq. 2.17 of \mich)

\eqn\sdc{\eqalign{& \langle V_T(k_1)
V_T(k_2)V_T(k_3)V_T(k_4)\rangle_{annulus} \sim  i \sqrt{G} G_o^4
(4\apm_{eff})^{-2}\delta^4 (k_1+k_2+k_3+k_4)\cr 
&\times \int_0^\infty 
{ds \over 2 \pi s^{11} }\eta({is \over \pi})^{-24} e^{- {\apm s \over
2}k_Ag^{AB}k_B}\cr
&\times \int_0^1d\n_1d\n_2d\n_3d\n_4  
\Psi_1\Psi_2\Psi_{12} e^{{i \over 2}\bigl[ k_3\times
k_4(2\n_{34}- \epsilon(\n_{34}))-k_1\times
k_2(2\n_{12}-\epsilon(\n_{12})) \bigr]},}} with
\eqn\wsx{\eqalign{\Psi_1&=|{\theta_{11}(\n_{12},{is \over \pi}) \over
\theta^\prime_{11}(0,{is \over \pi})}|^{2\apm_{eff} k_1 \cdot k_2}, ~~~~~
\Psi_2=|{\theta_{11}(\n_{34},{is \over \pi}) \over \theta^\prime_{11}(0,{is
\over \pi})}|^{2\apm_{eff} k_3 \cdot k_4},\cr
\Psi_{12}&=e^{-{s\over 4}}\prod_{r=1,2~s=3,4}|{\theta_{10}(\n_{rs},{is \over
\pi}) \over \theta^\prime_{11}(0,{is \over \pi})}|^{2\apm_{eff} k_r\cdot
k_s}, ~~~~~~~~ (\n_{rs}=\n_r-\n_s),
\cr k&=k_1+k_2, ~~~~~ k_r\times k_s =
k_{rA}\Theta^{AB}k_{sB}, ~~~~~~k_r\cdot k_s = k_{rA}G^{AB}k_{sB}.}}

The expression for the annulus amplitude in \sdc\ is written in the 
closed string channel. (The expression for the superstring would 
be similar except that
the factor of $ s^{-11} = s^{-d_t/2} \to s^{-3} $ in \sdc . This 
factor comes from the number of transverse dimensions $d_t$. )
Closed string singularities arise in the integral over $s$ in \sdc\ 
as $\eta({is \over \pi})$ may be expanded in a series in $e^{-Ns}$.
We thus find non analyticities\foot{
These singularites are 10 dimensional poles integrated over $d_t$ transverse
momenta.} (singularities) in the amplitude when
\eqn\rtfl{{\apm\over 2} k_Ag^{AB}k_B=-N .} 
In the NCOS
scaling limit, this condition may be written as 
\eqn\ght{{ \pi
\apm^2 \over \t G_o^4 }k_\m\eta^{\m\n}k_\n +{\t \over 4 \pi}
k_i\delta^{ij}k_j=-N.}  
Singularities on the real axis occur at a squared
energy $$k_0^2=k_1^2 + \left({G_0^2 \t \over 2 \pi \apm}\right)^2 
\left( k_2^2+k_3^2+{2N
\over \apm_{eff}} \right)$$ that becomes arbitrarily large as
$\apm$ is made increasingly small. In the strict limit $\apm \r
0$,  open string one loop amplitudes factorize on singularities of the
form $$ \int d^{d_t} k_M{ 1 \over k_2^2+k_3^2+{2N \over \apm_{eff}} 
+g^{MN}k_M k_N}.$$ 
As these singularities are never in the physical region, they do not
correspond to physical states.\foot{These singularities 
$\sim (k_i^2)^{{d_t -2 \over 2}} \ln (k_i^2)$ are very
similar to those induced by one loop graphs in spatially
noncommutative field theories, as found in \msv, \sv. 
Notice that if $d_t \ge 2$ ($p$ branes with $p<7$ in the supersymmetric
case), this amplitude, though non analytic,  is finite 
at $k=0$. For $d_t \leq 2$ the amplitude diverges at $k_i^2=0$.
It is possible that stronger IR singularities appear at higher
loops, specially for high dimensional branes. } 
Recalling that $G^{AB}$ is 
fixed in the NCOS limit, it is easy to 
see that the amplitude \sdc\ is finite (except of course for
the tachyon pole which is absent in the superstring). 
It is also straightforward using the
results of \mich\ to show that there are no physical poles for any
numbers of initial and final open string tachyon vertex operators.
Higher mass vertex operators involve additional powers of the 
Green functions on the annulus. These are finite in the NCOS limit and so
will not spoil the finiteness of the amplitudes.
Although we have not worked out the details, we expect that the 
behavior of the superstring is similar.

It is instructive to contrast the behaviour of \sdc\
in both the NCOS and the  NCYM limits. In the latter
case the $\apm \to 0$ limit is manifestly smooth 
when ${s \over \apm}$ is held fixed. This
forces one into a corner of the moduli space in which the massive open
string states are decoupled \refs{\andreev 
\kiem \gomis \mich \chu -\russo}. 
In the NCOS limit \sdc\ receives contributions
from finite $s$, and so from all open string oscillator states.  
Apart from the non-commutative phases 
the one loop open string diagram \sdc\ has almost the same form
as the corresponding diagram in a theory with $B=0$, with  
$\alpha'$ replaced by $\alpha'_{eff}$. 
However, the exponential term in \sdc\ coming from momentum flowing along the
closed string channel has a different $\alpha'$ dependence
from standard string theory with zero $B$. This different
dependence is responsible for the absence of physical closed string poles.

\fig{Nonplanar open string diagram. In open string field theory we would 
build it from the cubic vertex and we would consider states carrying 
momentum $q$ and $q+p$ along the loop.}{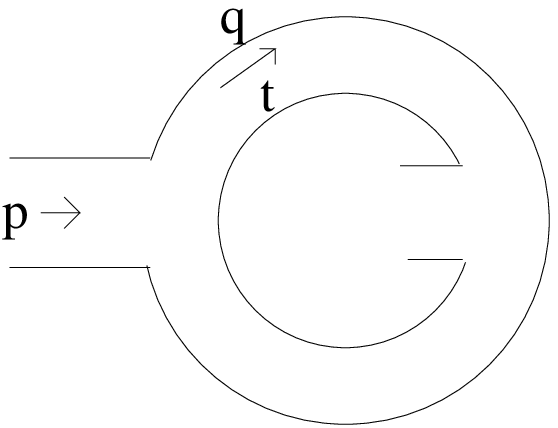}{5cm}

The absence of 
closed string poles in a non-commutative open string 
theory,  whose non-commutativity parameter $\theta$ is
 $2 \pi \alpha'_{eff}$ as in our NCOS theories, may be understood more 
directly, as we explain below.
This line of reasoning also suggests 
that a non commutative open string theory with $\t < 2 \pi \apm_{eff}$ 
has closed string poles, while the theory with  $\t > 2 \pi \apm_{eff}$
is unstable.
 
Consider the simple 
non-planar diagram represented in figure 1, in an open string field theory.
Let the open string theory in question be noncommutative, with 
noncommutativity parameter $\t$. The momentum integral 
for this diagram takes the form
\eqn\inte{
\int d^4 q  e^{ 2  i p \times q} I_{\theta=0}(q,p) \sim 
\int d^4 q \int_0^\infty dt e^{2  i p \times q }
 e^{ - 2\pi  \alpha'_{eff} t  q^2 +  t \beta p.q + ... }
}
where $I_{\theta=0}(p,q) $ 
is the integrand at $\theta =0$ and 
$p\times q = p_\mu q_\mu \Theta^{\mu\nu}/2$. 
We have exponentiated the propagators in the diagram using a Schwinger
proper time representation, where $t$ is the total proper time
along the loop and we have explicitly given the form of 
the leading dependence on $q$ ($\beta$ is some other Schwinger parameter, 
which is also integrated over; we have supressed this integral in 
\inte\ for simiplicity).
 When $q$ is integrated
over  we get  the diagram as a function of $t$ and $\b$.
As in \msv, the effect of noncommutativity on this integral is an 
extra term in the exponent of the form 
\eqn\extraterm{
e^{ - p\ o\ p/(8 \pi \alpha'_{eff} t). } ~
} 
where $p\ o \ p = - p_\mu \Theta^{\mu\nu}\Theta_{\nu\rho}p^\rho = - \theta^2
p^2$. 
This may be seen by shifting the 
integral over $q$ to one over $q'_\mu = q_\mu  + 
i \Theta_{\mu \nu} p^\nu/(4 \pi \alpha'_{eff}t)$. 
Note that terms of the form $q.p$ in \inte\ 
are unaffected by the shift due to the antisymmetry of $\Theta$.

Thus the integrand of \inte\ is modified from its 
$\theta =0$ value only by the additional exponential factor
\extraterm .
On shifting to the $s=\pi/t$ channel,  the integrand has the usual terms of 
the form 
$e^{ - s {\alpha'_{eff}\over 2} ( -p_0^2 + p_1^2 + ...) }$
(terms that  would produce the $s$-channel
poles if  $\theta$ were zero)  multiplied by the additional factor
$e^{-s{\t^2 \over 8 \pi^2 \apm_{eff} }(p_0^2 -p_1^2)}$.
When  $\theta = 2 \pi \alpha'_{eff}$ this extra factor exactly 
cancels the $p_0, p_1$ dependence of the exponent. Here we have used the
fact that we are in Lorenzian signature so that the final  sign of
the exponent in \extraterm\ is the opposite to the one in 
 Euclidean signature. 
If $\theta$ is  slightly less that its critical value, then 
\extraterm\ 
does not  cancel the closed string poles. If $\theta$ is bigger
than its critical value then all closed string poles 
turn tachyonic, a reflection of the instability of the system.

\newsec{Higher Loop diagrams} 

In this section we will examine higher loop string diagrams in the 
NCOS limit. We will not attempt to prove that the limit is nonsingular 
for arbitrary diagrams, but 
we will observe that a simple counting of powers of $\apm$ does not 
reveal any difficulties.   
Naively, a genus $g$ surface in the string loop expansion 
is weighted by $g_{str}^{2g-2}$. As $g_{str}$ diverges in the NCOS
limit, a perturbative expansion in genus seems impossible.
However, we shall argue below that both holes and handles
are really weighted by powers of $G_0$ and so high genus surfaces 
are suppressed at weak open string coupling.

\subsec{Holes}

The addition of a hole in the world sheet is accompanied by one 
power of $g_{str}$. It also leads to an additional integral over the zero mode
momentum circulating around the loop. As shown in \call, these 
integrals have a measure factor proportional to 
${\rm det}^{1/2}(g+2\pi \apm B){\rm det}^{-1/2}(g)$. 
Hence the total weighting of a hole is 
\eqn\hwt{g_{str}{ {\rm det}^{1/2}(g+2\pi \apm B)
\over {\rm det}^{1/2}(g)}=G_o^2,}
and is finite as $\apm \to 0$.

\subsec{Handles}

Consider an open string world sheet $A$, with open string boundary 
conditions corresponding to a 3-brane. The amplitude on a worldsheet $(B)$
with an additional handle can be factorized in the closed string channel
along the handle. The resultant amplitude reads schematically as 
$$S_B= \sum S_{A_{V_a, V_a}} \l_{eff}^2 
\int d^6k {1 \over g^{IJ}k_{I}k_{J}+m_a^2}; ~~~~~~(I,J=0\ldots 9).$$
Here $S_{A_{V_a, V_a}}$ denotes the amplitude on $A$ with two extra closed
string insertions. The integral is over the momenta of the intermediate 
states in the transverse directions (momentum 
is not conserved in these directions).

\fig{Adding a handle to a worldsheet $A$, we obtain a worldsheet $B$, which
can be represented as coming from the propagation of closed string states
between two points  of the worldsheet. We sum over all closed string
states.}{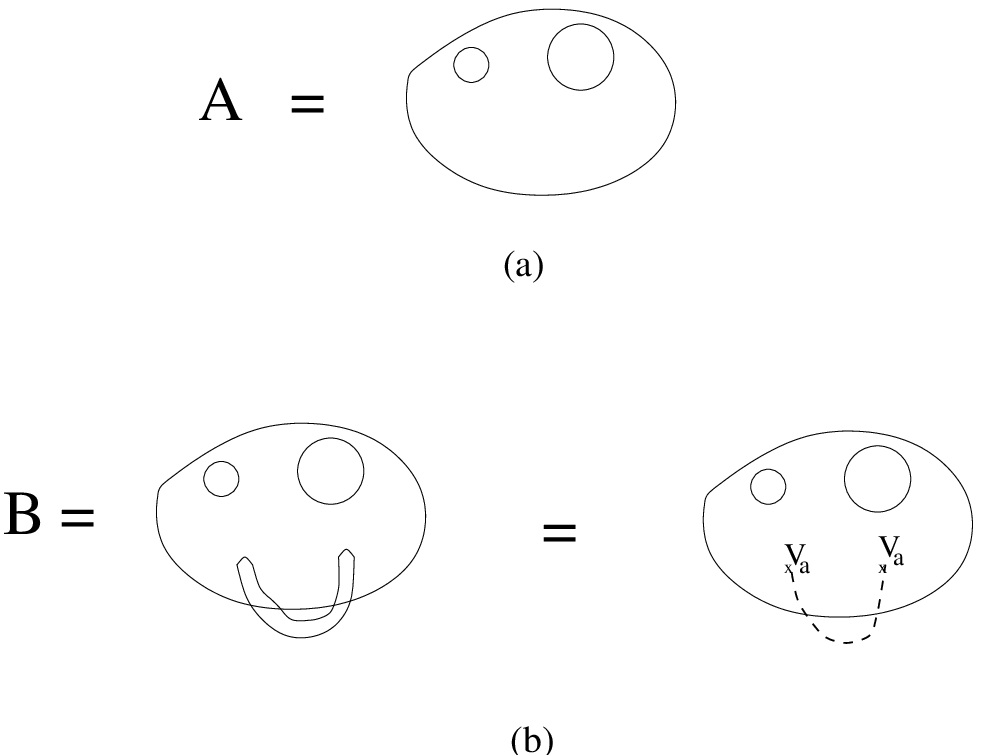}{7cm}

\noindent The effective coupling $\l_{eff}^2$ is determined as follows: 
A closed string mode $\ph$ with spacetime action 
$$S={ 1 \over g_{str}^2 \apm^4} \int d^{10}x \sqrt{g} 
(\p_{I} \ph \p_{J} \ph g^{IJ}+ m_a^2 \ph^2)$$
has effective coupling 
$$\l_{eff}={g_{str} \apm^2 \over g^{{1\over
4}}} ={ \apm^{{ 5 \over 2}}  \over G_0^4 \apm_{eff}^{{\half}} }$$
in the NCOS limit. 
The integral 
$$\int d^6k {1 \over g^{IJ}k_{I}k_{J}+m_a^2}=
\apm \int d^6k {1 \over { \apm ^2 k_Mk_N \d^{MN}\over \apm_{eff}G_0^4}
+ N +\apm_{eff} (k_2^2 +k_3^2) + ...}$$
is of order
$${ \apm} \left( {\apm_{eff} G_0^4 \over \apm^2} \right)^3.$$
Finally, in the normalization we have adopted, 
$S_{A_{V_a, V_a}}$ is of the same order as $S_A$. 

Putting it all together, we find that 
\eqn\handlewt{{ S_B  \over S_A} \approx { G_0^4 \apm_{eff}^2}.}
Thus we conclude that extra handles, in the NCOS limit, are 
neither infinitely suppressed nor enhanced in the NCOS limit.
They are instead really weighted by a 
factor of $G_0^4$, as they would have been 
for an ordinary weakly coupled open string theory\foot{See also
\oleg\ for a discussion of diagrams with many holes.}.

\newsec{Supergravity duals}

The considerations of the previous sections generalize to open string 
theories on $N$ coincident 3-branes. 
In that case since we are dealing with a deformation of $U(N)$ ${\cal N}=4$ SYM
we expect that it should have a supergravity dual for large $N$. 
The relevant  supergravity solutions were written in  \refs{\ih,\mr}. 
 We start from 
the Lorentzian version of the  solution (2.3) in \mr , with $B_{23} =0$.
 Then we do the
following scaling of parameters
\eqn\scaling{\eqalign{
r = & \sqrt{\alpha'} u 
\cr
\cosh \theta' = & {\tilde b' \over \alpha'}
\cr
g =& { \tilde g  \tilde b' \over \alpha'}
\cr
x_{0,1} =& { {\tilde b'}  \over \sqrt{\alpha'}} \tilde x_{0,1}
\cr
x_{2,3} = & \sqrt{\alpha'} \tilde x_{2,3}
\cr
R^4 =& fixed = { 4 \pi \tilde g }N  
}}
 We obtain\foot{Here we normalize the $B$ field as in the previous
sections,
in \mr\ it was normalized differently by a factor of $2 \pi \alpha'$,
$B_{MR} = 2 \pi \alpha' B_{here} $.} 
\eqn\declor{\eqalign{
ds^2_{str} &=
 \alpha ' f^{1/2}  \left[ { u^4 \over  R^4 }
( - d{\tilde x_0}^2 +d{\tilde x_1}^2 ) + f^{-1} 
( d{\tilde x_2}^2 +d{\tilde x_3}^2 ) 
 + du^2 + u^2 d \Omega^2_5 \right]
\cr
2 \pi \alpha' B_{01} &= \alpha' 
 { u^4 \over R^4 } \ ,\ \ \ \ \ \
\cr
e^{2\phi} &= { {\tilde g}^2  } f {u^4 \over R^4}
\cr
A_{23} = & \alpha'
 { 1 \over \tilde g} f^{-1}\ ,
\cr
F_{0123u} & = {\alpha'}^2
{1 \over \tilde g}   {4 f^{-1} \over u }
\cr 
f & = 1 + { R^4 \over u^4 } 
}}                                                        
The particular scalings that we have to do to reproduce this solution 
are, up to constants,  the same as those  in section 2.2. 
The only scaling that is not so obvious  is the scaling of the radial 
coordinate. Notice that in the ${\cal N } = 4$ SYM case we rescale the
radial coordinate as $ r \sim \alpha' u$. The fact that we have 
$r \sim \sqrt{\alpha'} u $ in this case is related to the fact that
the closed string metric has a factor of $1/\alpha'$ in section 2.2.
We see from 5.2  that for small $u$ we recover the usual 
$AdS_5\times S^5$ solution as we expect, since the open string theory
reduces to ${\cal N}=4$ SYM at low energies. In particular we see
that we should identify $\tilde g = G_0^2$. As we increase $u$ 
the metric becomes different than the metric of $AdS$ and
we also see that  the dilaton becomes large. 
This suggests that for large $u$ we should do an  S-duality 
to analyze the solution. After we do the S-duality we obtain
a solution which is the same as the supergravity solution which 
corresponds to a D3 brane with spatial non-commutativity in the
directions 23, see \mr , eqn. (2.7). 
 This suggests that at very high energies
the open string theory we are  studying would have a dual description
in terms of the theory with spatial non-commutativity. 

\newsec{Discussion}

\subsec{Open String Dipoles and UV/IR}

Free open string states in the NCOS limit behave quite 
differently from ordinary open strings propagating in the same metric, despite
having the same spectrum.
In the presence of background fields, 
(as discussed for example in \refs{\burg,\bach,\nest} and especially in 
\mich\ for the magnetic case) 
the mode expansion reads
\eqn\faxz{X^\m(\sigma, \tau)=x_0^\m+2i\apm_{eff}p^\m \tau +
{1 \over \pi}\Theta^{\m\n}p_\n\sigma +({\rm oscillators}).}
For strings in the NCOS limit this
implies that the distance along the direction of the field 
between the two ends of the string , as measured in the metric $G^{AB}$, is 
\eqn\rfkz{\Delta X^1= 2\pi\apm_{eff}k_0,}
plus oscillator contributions which time average to zero. 
(Note that $\D X^1$ is the distance between the endpoints of the 
string worldsheet along a line of constant worldsheet time 
rather than along a line of constant $X^0$. As we argue below, 
the proper length of the string is given by a formula analogeous 
to \rfkz\ with $k_0$ replaced by the centre of mass energy of the 
state.)

The invariant energy and proper length of an oscillator state may 
be estimated as follows. The 
tension of an open string aligned with a near critical electric field is  
almost canceled by a negative contribution from its dipole interaction
with the field. In the NCOS limit, the effective tension $T_{eff}={1
\over 4 \pi \apm_{eff}}$ (see \epul) . The energy of such a string,
with an oscillation number $N$, is $E = T_{eff} L +{\pi N\over L}$. 
This is 
minimized for $L = 2 \pi  \sqrt{\alpha'_{eff} N} = 
 2 \pi \alpha'_{eff} E $, as in \rfkz .

\subsec{Thermodynamics}

At low energies, compared to ${1 \over \sqrt{\apm_{eff}}}$ 
the NCOS theory reduces to ordinary $\CN=4$ SYM, and its 
free energy scales like $T^4$. At intermediate energies, 
the thermodynamics of a weakly coupled NCOS theory 
($\l=G_o^2N  \ll 1$), may be expected to reflect its Hagedorn 
density of states.

However, as argued in this paper, the weakly coupled NCOS theory 
has a dual description as a strongly coupled NCYM theory. 
In a spatially 
noncommutative field theory, at weak coupling, planar diagrams
\rfilk\
dominate over nonplanar diagrams \msv\ for energies $k_0 \gg {1 \over
\sqrt{\t}}$. It is plausible that this result to continues to
hold at strong coupling\foot{This statement is true  
at least in the `supergravity' limit $\l \gg 1$,
$G_o^2 \ll 1$; in that limit \mr, \ih, 
supergravity suggests that planar diagrams
dominate for $k_0 \gg {1 \over (\l \t^2)^{{1 \over 4}}}$.}, 
with a crossover scale renormalized
by a function of the coupling. If true, this assertion implies that, 
at high temperatures,  
the free energy of spatially noncommutative 
SYM is proportional to the free energy of ordinary large $N$
SYM, and so scales with temperature like $T^4$, even at large $G_o^{'2}$.

It would be interesting to investigate this issue further.

\subsec{Generalizations to other Dimensions}

In this paper we have `derived' the existence of a decoupled 
four dimensional open string theory, NCOS, by $S$ dualizing spatially 
noncommutative SYM. We presented evidence that, independent of this
derivation, the resultant theory is well defined, and weakly coupled
over a range of parameters.  

It is easy to extend our construction of the NCOS to other dimensions,
even though we do not have an independent ($S$ duality) 
argument for the decoupling of closed strings.
The NCOS scaling limit for a $p$ brane is, once again,  defined by  table 1, 
where the indices $i,j$ run from $2\ldots p$
and $A,B$ from $0\ldots p$. In other words, this limit still 
describes a near critical electric field turned on in the $1$
direction. The open string coupling defined in \gya\ and 
the effective low energy Yang Mills coupling constant 
$g_{YM}^2 \sim G_0^2\apm_{eff}^{{p-3 \over 2}}$ are finite.
In the NCOS limit, open strings  appear to decouple 
from closed strings for all $p$. 
The annulus amplitude is finite in arbitrary dimension, 
and always factorizes on unphysical closed string poles.
As in the 3-brane, string diagrams with handles and holes are  suppressed
by powers of the open string coupling, and may be neglected at weak
coupling. 

In fact, these open string theories appear to be 
non-gravitational UV finite completions of low energy  
(supersymmetric) Yang-Mills. 
This statement appears to be true even in high dimensions 
where  the gauge theory is non-renormalizable.

\centerline{\bf Acknowledgements}
We are grateful to C. Bachas, M. Berkooz, M. Gutperle, J. Harvey, S. Kachru,
H. Liu, J. McGreevy, P. Kraus, N. Seiberg, A. Sen, S. Shenker, 
E. Silverstein, L. Susskind and N. Toumbas
for useful discussions. R.G. and S.M. are grateful to
the high energy theory group at Stanford
University for their hospitality.
 
This work was supported in part by DOE grant DE-FG02-91ER40654.

\listrefs
\end